\documentclass[12pt]{article}

\usepackage[T1]{fontenc}
\usepackage{calc}
\usepackage{setspace}
\usepackage[normalem]{ulem}
\usepackage{graphicx}
 
\begin{document}
\begin{spacing}{1.66}
{\large \bf Magnetotransport in a bi-crystal film of La$_{\rm{}0.7}$Sr$_{\rm{}0.3}$MnO$_{\rm{}3}$}
\end{spacing}

\begin{spacing}{1.66}
R. Mathieu$^{\rm{}a}$, P. Svedlindh$^{\rm{}a}$, R. Chakalov$^{\rm{}b}$, and Z. G. Ivanov$^{\rm{}b}$.
\end{spacing}

\begin{spacing}{0.66}
{\it \small $^{\rm{}a}$ Department of Materials Science, Uppsala University, Box 534, SE751 21 Uppsala, Sweden}
\end{spacing}

\begin{spacing}{1.66}
{\it  \small $^{\rm{}b}$ Department of Physics, Chalmers University of Technology, SE412 96 Göteborg, Sweden}
\end{spacing}

\begin{spacing}{1.66}

\end{spacing}

\begin{spacing}{1.66}
Transport properties of an epitaxial film of La$_{\rm{}0.7}$Sr$_{\rm{}0.3}$MnO$_{\rm{}3}$ (LSMO), deposited epitaxially on a LaAlO$_{\rm{}3}$ bi-crystal substrate having a misorientation angle of 9.2$^{\rm{}o}$, have been studied. The film was patterned into a meander containing 100 grain boundaries. The resistivity of the sample exhibits two components; one originating from the grain boundary regions, and one from the LSMO elements in the meander; the latter contribution is similar to the resistivity of a reference epitaxial LSMO film. The low (<0.5 T) and high (up to 6 T) field magnetoresistance was also studied. The meander show a large low field magnetoresistance, increasing with decreasing temperature, and a constant high field slope of the magnetoconductance, results that are well explained by a two-step spin polarized tunneling model.
\end{spacing}

\begin{spacing}{1.66}

\end{spacing}

\begin{spacing}{1.66}
\end{spacing}

\begin{spacing}{1.66}

\end{spacing}

\begin{spacing}{1}
{\it \small Corresponding author: Roland Mathieu, FTF, Box 534, SE751 21 Uppsala, Sweden,\\ \hspace*{4.4cm} Fax: +46 18 500 131, Email: roland.mathieu@angstrom.uu.se}
\end{spacing}

\begin{spacing}{1.66}

\newpage

Single crystals [1] and epitaxial films [2] of manganites show a colossal magnetoresistance (CMR), but only for relatively large magnetic fields and close to the ferromagnetic transition temperature (\textit{T}$_{\rm{}c}$). Polycrystalline films [1] instead, as well as bi-epitaxial films [3] and films grown on bi-crystal substrates [4], exhibit at temperatures \textit{T}<<\textit{T}$_{\rm{}c}$ a large low field response.
\end{spacing}

\begin{spacing}{1.66}
A La$_{\rm{}0.7}$Sr$_{\rm{}0.3}$MnO$_{\rm{}3}$ (LSMO) thin film was fabricated by depositing LSMO (a$_{\rm{}LSMO}$=3.82 Å) epitaxially on a LaAlO$_{\rm{}3}$ (LAO) bi-crystal substrate (a$_{\rm{}LAO}$=3.79 Å); the film obtained was then patterned into a meander containing one hundred 18.4$^{\rm{}o}$(9.2/9.2) oriented grain boundary links. Measurements were either made on all 100 junctions (100J) or on one junction (1J). Structural properties were checked with Xray $\theta$-2$\theta$- and $\phi$-scans. The crystalline quality of the meander is, apart from the grain boundaries, similar to that of a reference epitaxial LSMO film (EF). The resistivity \textit{$\rho$}(\textit{H},\textit{T},\textit{$\theta$}) was measured using a Maglab 2000 system equipped with a rotationary probe. The magnetoresistance of the samples is defined as MR=(\textit{R}$_{\rm{}0}$-\textit{R}$_{\rm{}H}$(\textit{$\theta$}))/\textit{R}$_{\rm{}0}$; the angle \textit{$\theta$} refers to the angle between the inplane magnetic field and the current.
\end{spacing}

\begin{spacing}{1.66}
The zero field resistivity for 1J and 100J was recorded vs. temperature. The overall behavior is similar to that of the EF film; for 1J an additional 'knee' is present at \textit{T}$\approx$300 K (\textit{T}$_{\rm{}c}$$\approx$360 K) revealing a contribution to the measured resistivity originating from the grain boundary. This contribution is less apparent in the results for 100J, indicating that there is some scatter in the properties of the different junctions.
\end{spacing}

\begin{spacing}{1.66}
Fig. 1 shows the temperature dependence of the low field MR(\textit{H}=0.1 T) for all samples; for 1J and 100J results are shown for different magnetic field orientations. EF displays the typical low field magnetoresistance behavior for a high quality epitaxial film [5] with a peak in the magnetoresistance around \textit{T}$_{\rm{}c}$ and no significant low temperature MR. In comparison, the results for the meander show a low temperature tail, larger when the current and field are parallel. MR(\textit{T}=77K) $\approx$ 6\% in the (\textit{H} $\|$ \textit{I}) case, which is in agreement with earlier results on similar bi-crystal films [6];  At higher temperature, the behavior is rather close to that of EF. One may notice that due to shape anisotropy of the meander, the domain magnetization is in plane, perpendicular to a grain boundary. This is evidenced by results obtained for the low field magnetoresistance (cf. Fig. 2); applying the field along the current, the magnetoresistance shows two hysteretic peaks at fields close to the coercive field of the EF sample. Applying the field perpendicular to the current (along a grain boundary), a reversible magnetoresistance is detected, indicating that the magnetization process in this case corresponds to reversible rotation of the domain magnetization. 
\end{spacing}

\begin{spacing}{1.66}
As first proposed by Lee et al. [7], for polycrystals, and as observed in bi-epitaxial grain boundary LSMO films [3], the high field magnetoconductance rather than the magnetoresistance is linear with field. This behavior is consistent with a model based on secondorder spin polarized tunneling through interfacial spin sites. Using the transfer integral $T_{12}\approx \sqrt{1+\uline{s}_1\uline{s}_2}$ for itinerant e$_{\rm{}g}$ electrons between localized t$_{\rm{}2g}$ moments (\textit{\uline{s}}$_{\rm{}i}$ is the normalized spin moment), the grain boundary conductivity is given as \textit{G}$_{\rm{}j}$$\sim{}$\textit{T}$_{\rm{}1j}$$^{\rm{}2}$\textit{T}$_{\rm{}j2}$$^{\rm{}2}$, which becomes \textit{G}$_{\rm{}j}$$\sim${}<\textit{\uline{s}} $_{\rm{}j}$>$\sim{}${}\textit{$\chi$}$_{\rm{}j}$\textit{H} for large \textit{H}. Here <\textit{\uline{s}} $_{\rm{}j}$> is the thermal average of the grain boundary moment and \textit{$\chi$}$_{\rm{}j}$ is the corresponding susceptibility. The temperature dependence of the normalized high field slope of the magnetoconductance \textit{b}=($\mu_{\rm{}0}$\textit{G}$_{\rm{}0}$)$^{\rm{}{-1}}$ $\partial$\textit{G}/$\partial$\textit{H}$\sim${}\textit{$\chi$}$_{\rm{}j}$ is shown in Fig. 3; the results obtained for a bi-epitaxial grain boundary LSMO film [3] (GBF) are included for comparison. The results for GBF are dominated by properties of the grain boundaries, while the results for the meander contain contributions both from grain boundaries and epitaxial LSMO elements. The contribution from the grain boundary is particularly evident in the 1J results; the maximum in \textit{b} occurs at the same temperature as where we observe a 'knee' in the resistance vs. temperature curve.
\end{spacing}

\begin{spacing}{1.66}

\end{spacing}

\begin{spacing}{1.66}
Financial support from NFR is gratefully acknowledged.
\end{spacing}

\begin{spacing}{1.66}

\end{spacing}

\begin{spacing}{1.66}
[1] H. Y. Hwang, SW. Cheong, N. P. Ong, and B. Batlogg, Phys. Rev. Lett. 77 (1996) 2041.
\end{spacing}

\begin{spacing}{1.66}
[2] M. F. Hundley, M. Hawley, R. H. Heffner, Q. X. Jia, J. J. Neumeier, J. Tesmer, J. D. Thompson, and X. D. Wu, Appl. Phys. Lett. 67 (1995) 860.
\end{spacing}

\begin{spacing}{1.66}
[3] R. Mathieu, P. Svedlindh, R. Chakalov and Z. G. Ivanov, Phys. Rev. B 62 (2000) 3333.
\end{spacing}

\begin{spacing}{1.66}
[4] K. Steenbeck, T. Eick, K. Kirsch, K. O'Donnell, and E. Steinbei, Appl. Phys. Lett. 71 (1997) 968.
\end{spacing}

\begin{spacing}{1.66}
[5] J. E. Evetts, M. G. Blamire, N. D. Mathur, S. P. Isaac, B. S. Teo, L. F. Cohen, and J. L. MacmanusDriscoll, Phil. Trans. Soc. Lond. A 356 (1998) 1593.
\end{spacing}

\begin{spacing}{1.66}
[6] S. P. Isaac, N. D. Mathur, J. E. Evetts, and M. G. Blamire, Appl. Phys. Lett. 72 (1998) 2038.
\end{spacing}

\begin{spacing}{1.66}
[7] S. Lee, H, Y. Hwang, B. I. Shraiman, W. D. Ratcliff II, and SW Cheong, Phys. Rev. Lett. 82 (1999) 4508.
\end{spacing}

\begin{spacing}{1.66}

\end{spacing}

\begin{spacing}{1.66}
\newpage
\begin{figure}[h]
\begin{center}\leavevmode
\includegraphics[width=0.5\linewidth]{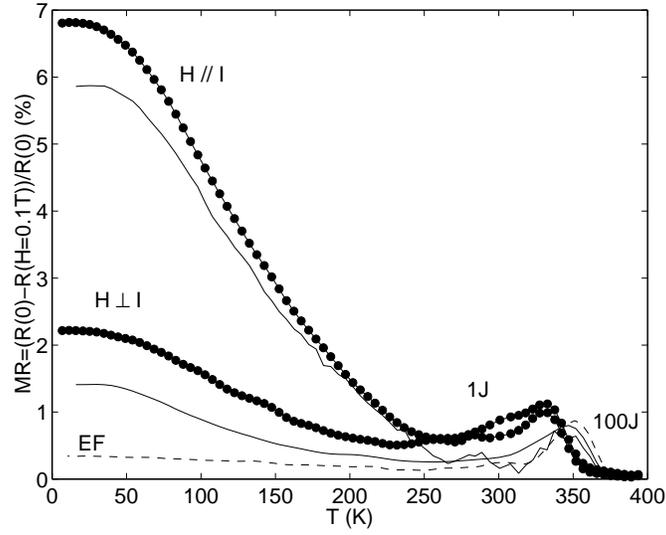}
\caption{Temperature dependence of the low field (\textit{H}=0.1T) magnetoresistance of the bi-crystal (BC) film for different magnetic field orientations; results for 1 (1J, filled circles) and 100 (100J, line) junctions are included. The magnetoresistance of a high quality exitaxial film (EF, dotted line) is added for comparison.}
\end{center}
\end{figure}
\end{spacing}

\begin{spacing}{1.66}

\end{spacing}

\begin{spacing}{1.66}
\begin{figure}[h]
\begin{center}\leavevmode
\includegraphics[width=0.5\linewidth]{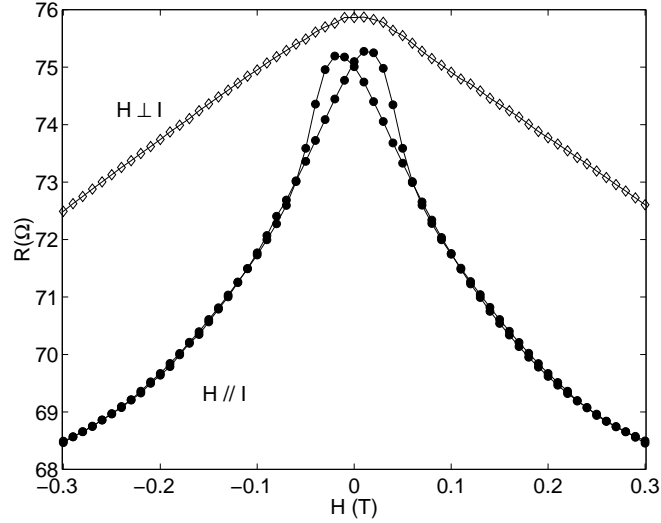}
\caption{\textit{R}(\textit{H}) for different magnetic field orientations; \textit{T}=80K. In the (\textit{H} $\|$ \textit{I}) case, the magnetic field is parallel to the applied current, and thus perpendicular to the grain boundary; it is the opposite in the (\textit{H} $\perp$ \textit{I}) case.}
\end{center}
\end{figure}
\end{spacing}

\begin{spacing}{1.66}

\end{spacing}

\begin{spacing}{1.66}
\begin{figure}[h]
\begin{center}\leavevmode
\includegraphics[width=0.5\linewidth]{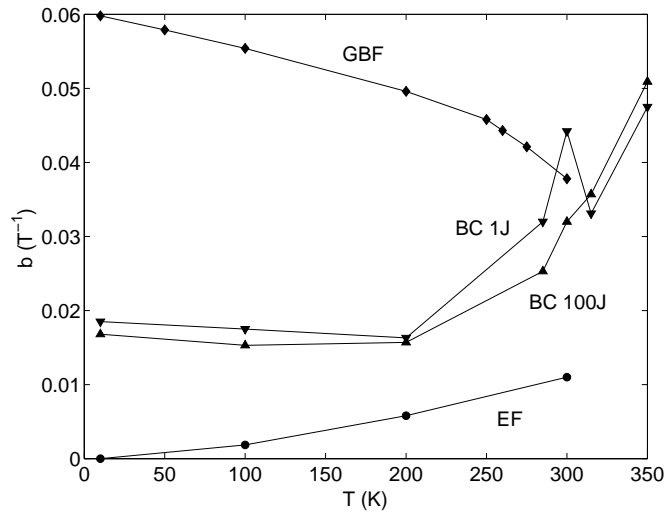}
\caption{Temperature dependence of the normalized high field magnetoconductance slope \textit{b} for the bi-crystal film (BC) for 1 and 100J; results obtained for a bi-epitaxial film (GBF) [3] and a high quality epitaxial film (EF) are included for comparison.}
\end{center}
\end{figure}
\end{spacing}

\begin{spacing}{1.66}

\end{spacing}

\begin{spacing}{1.66}

\end{spacing}

\begin{spacing}{1.66}
\end{spacing}

\end{document}